\documentclass[runningheads]{llncs}
\usepackage{amssymb}
\usepackage{graphicx}
\usepackage{llncsdoc}
\usepackage{makeidx, epsfig,amsmath}  
\usepackage{url}

\usepackage[colorinlistoftodos]{todonotes}
\usepackage[colorlinks=true, allcolors=blue]{hyperref}
\usepackage{caption}
\usepackage[title]{appendix}
\usepackage{amsfonts}
\usepackage{multirow}
\usepackage{booktabs}
\usepackage[ruled,vlined,algo2e]{algorithm2e}

\usepackage{epsfig}
\usepackage{graphicx}
\usepackage{amsmath}
\usepackage{amssymb}
\usepackage{algorithm}
\usepackage{algorithmic}
\usepackage{subfigure}
\usepackage{caption}
\usepackage{epsfig}
\usepackage{epstopdf}
\usepackage{wrapfig}
\usepackage{multirow}
\usepackage{array}
\usepackage{tabulary}
\usepackage{graphicx}
\usepackage{color}

\newcolumntype{K}[1]{>{\centering\arraybackslash}p{#1}}

\begin{document}
%
%

\title{DeepPrognosis: Preoperative Prediction of Pancreatic Cancer Survival and Surgical Margin via Contrast-Enhanced CT Imaging}

\author{Jiawen Yao\inst{1} \and
Yu Shi\inst{2} \and
Le Lu\inst{1} \and
Jing Xiao\inst{3} \and
Ling Zhang\inst{1}}
\authorrunning{J. Yao et al.}
\titlerunning{DeepPrognosis: Preoperative Prediction of Pancreatic Cancer Survival}

\institute{PAII Inc., Bethesda, MD 20817, USA \and
Shengjing Hospital of China Medical University, Shenyang, PR China \and Ping An Technology, Shenzhen, PR China}
\maketitle              

\begin{abstract}
Pancreatic ductal adenocarcinoma (PDAC) is one of the most lethal cancers and carries a dismal prognosis. Surgery remains the best chance of a potential cure for patients who are eligible for initial resection of PDAC. However, outcomes vary significantly even among the resected patients of the same stage and received similar treatments. Accurate preoperative prognosis of resectable PDACs for personalized treatment is thus highly desired.
Nevertheless, there are no automated methods yet to fully exploit the contrast-enhanced computed tomography (CE-CT) imaging for PDAC. Tumor attenuation changes across different CT phases can reflect the tumor internal stromal fractions and vascularization of individual tumors that may impact the clinical outcomes. In this work, we propose a novel deep neural network for the survival prediction of resectable PDAC patients, named as 3D Contrast-Enhanced Convolutional Long Short-Term Memory network (CE-ConvLSTM), which can derive the tumor attenuation signatures or patterns from CE-CT imaging studies. 
We present a multi-task CNN to accomplish both tasks of outcome and margin prediction where the network benefits from learning the tumor resection margin related features to improve survival prediction. The proposed framework can improve the prediction performances compared with existing state-of-the-art survival analysis approaches. The tumor signature built from our model has evidently added values to be combined with the existing clinical staging system.
\end{abstract}


\section{Introduction}
\noindent 
Pancreatic ductal adenocarcinoma (PDAC) is one of the most lethal of all human cancers, which has an extremely very poor 5-year survival rate of 9\%~\cite{siegel2019cancer}. Surgical resection, in combination with neoadjuvant chemotherapy, is the only potentially curative treatment for PDAC patients. Offering surgery to those who would most likely benefit (e.g., a high chance of long-term survival) is thus very important for improving life expectancy. 
Computed tomography (CT) remains the primary initial imaging modality of choice for the pancreatic cancer diagnosis. Previous work adopts image texture analysis for the survival prediction of PDACs~\cite{attiyeh2018survival}. However, the representation power of hand-crafted features on only venous phase CT might be limited. More recently, deep learning based approaches have shown good performances not only in traditional medical imaging diagnosis~\cite{wang2017chestx,wang2019weakly}, but also in prognosis models like outcome prediction of lung cancer~\cite{lou2019image,Yao2017DeepCorr,zhu2017wsisa} and gliomas~\cite{nie20163d,liu20193d}. The success of 3DCNNs contributes to capture deep features not only in the 3D gross tumor volume but also in peritumoral regions. However, such models may not generalize well for PDAC because some important predictive information is not necessarily existing in isolated imaging modality/phase.
Contrast-enhanced computed tomography (CE-CT) imaging plays a major role in the depiction, staging, and resectability evaluations of PDAC. Typical characteristics of pancreatic cancer include a hypo-attenuating mass during the pancreatic and venous phases observed in CE-CT. Enhancement variations among multiple CT phases are known to reflect the differences in internal stromal fractions and vascularization of individual PDACs that impact clinical outcomes~\cite{prokesch2002isoattenuating}. However, the use of multi-phase CE-CT for quantitative analysis of PDAC prognosis has not been well investigated in the literature. There is also no established deep learning survival models 
to incorporate enhancement variations in multi-phase CE-CT. The margin resection status, which is known to be associated with the overall survival (OS) of PDAC patients~\cite{konstantinidis2013pancreatic}, has been incorporated into the previous radiomics-based PDAC prognosis model~\cite{attiyeh2018survival}. However, the margin status is only available after the surgery is conducted. A preoperative prediction of the margin resection is desired but has not been investigated.  
\begin{figure*}[!htb]
	\centering
	\includegraphics[width=0.8\linewidth]{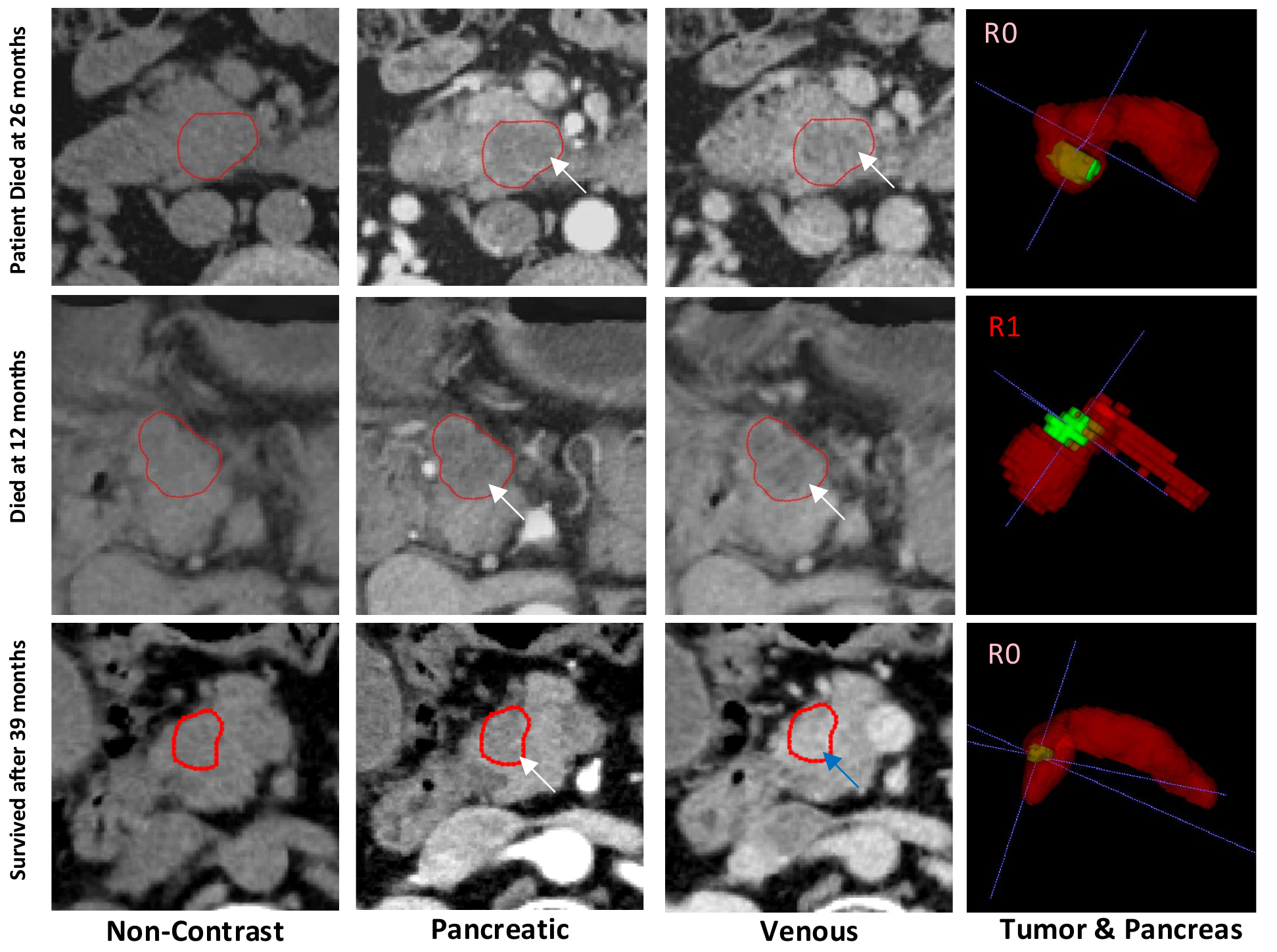}
	\caption{Example of the multi-phase CE-CT images and PDAC tumor enhancement patterns. Tumor and pancreas masks show the spatial relationship between the pancreas and tumor. The white arrow depicts a hypo-attenuating tumor; blue arrow indicates an iso-attenuating mass.}
	\label{fig: panc_vis} 
\end{figure*}

In this paper, we propose a novel 3D Contrast-Enhanced Convolutional Long Short-Term Memory (CE-ConvLSTM) network to learn the enhancement dynamics of tumor attenuation from multi-phase CE-CT images. This model can capture the tumor's temporal changes across several phases more effectively than the early fusion of input images. Furthermore, to allow the tumor resection margin information to contribute to the survival prediction preoperatively,
we present a multi-task learning framework to conduct the joint prediction of margin status and outcome. The jointly learning of tumor resectability and tumor attenuation in a multi-task setting can benefit both tasks and derive more effective/comprehensive prognosis related deep image features. Extensive experimental results verify the effectiveness of our presented framework. The signature built from the proposed model remains strong in multivariable analysis adjusting for established clinical predictors, and can be combined with the established criteria for risk stratification and management of PDAC patients.

\section{Methods}

The preoperative multi-phase CE-CT pancreatic imaging used in this study have been scanned at three time points for PDACs located at the pancreas head and uncinate. After the non-contrast phase, average imaging time delays are 40-50 seconds for the pancreatic phase and 65-70 seconds for the portal venous phase. Fig.\ref{fig: panc_vis} shows three examples to illustrate different tumor attenuation and resection margins of PDAC patients. Tumor attenuation in specific CT phases are very important characteristics to identify and detect the tumor. Each row in Fig.\ref{fig: panc_vis} represents one PDAC patient, and red boundaries are the tumor annotations. The white arrow indicates a typical hypo-attenuating tumor, while blue arrow shows an iso-attenuating tumor. In previous studies, Kim et al. reported that visually iso-attenuating PDACs are associated with better survival rates after surgery, as opposed to typical hypo-attenuating PDACs~\cite{kim2010visually}. 
Hypo-attenuating mass can be clearly observed in both pancreatic and venous phases of the first and second patients, indicating low stromal fractions (worse clinical outcomes). For the third patient, even though tumor hypo-attenuating is observed in pancreatic phase, it then reflects iso- or even hyper-attenuating in the venous phase compared with its adjacent pancreas regions, indicating high stromal fractions (better survival). This reminds us that tumor enhancement changes across phases is a very useful marker to reflect tumor internal variations and could benefit prognosis.
Besides tumor attenuation, another very important factor is the resection margin indicating the margin of apparently non-tumorous tissue around a tumor that has been surgically removed. 
More specifically, the resection margin is characterized as R0 when no evidence of malignant glands was identified microscopically at the primary tumour site. R1 resections have malignant glands infiltrating at least one of the resection margins on the permanent section and are usually associated with poor overall survival~\cite{konstantinidis2013pancreatic}. From the Fig.\ref{fig: panc_vis}, both tumors from the first and second patient display hypo-attenuating appearances, but it is clear to see that the second tumor has infiltrated out of the pancreas shown in tumor and pancreas masks. The pathological evidence indicates the second patient has the PDAC with R1 resections, and a follow-up study shows this patient has worse outcome than the first patient. Radiological observations about tumor attenuation and surgical margins status from CE-CT imaging motivate us to develop a preoperative PDAC survival model.  

\begin{figure*}[t]
	\centering
	\includegraphics[width=1.0\linewidth]{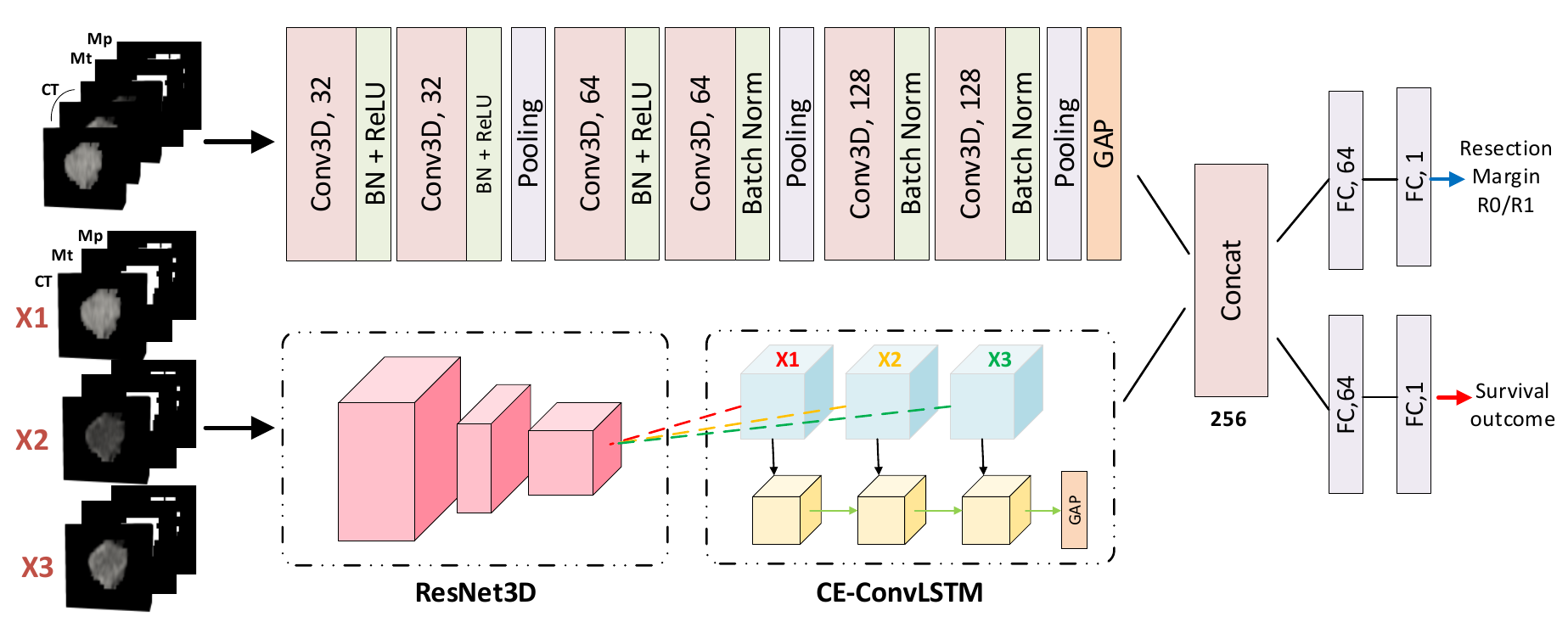}
	\caption{An overview of the proposed multi-task model with CE-ConvLSTM.}
	\label{fig: Framework} 
\end{figure*}

We use time points 1, 2, 3 to represent non-contrast, pancreatic, and venous phases, respectively. A radiologist with 18 years of experience in pancreatic imaging manually delineates the tumors slice by slice on the pancreatic phase. The segmentation of the pancreas is performed automatically by a nnUNet model \cite{isensee2018nnu} trained on a public pancreatic cancer dataset with annotations~\cite{simpson2019large}. We use DEEDS \cite{heinrich2013towards} to register the non-contrast and venous phases to the pancreatic phase. Three image feature channels are derived: 1) CT-image of pancreas and PDAC (background removed) in the soft-tissue window [-100, 200HU] and are normalized as zero mean and unit variance; 2) binary tumor segmentation mask and 3) binary pancreas segmentation mask. A multi-phase sequence of image subvolumes of $64\times64\times64$ pixels$^{3}$ centered at the tumor 3D centroid are cropped to cover the entire tumor and its surrounding pancreas regions. The dataset is prepared for every tumor volume from each phase scan, to build the 4D CE-CT data for both training and testing (as $X_t = \{X_t^{CT}, X_t^{M_T}, X_t^{M_P}\}, t \in \{1,2,3\}$, $M_T$: tumor mask, $M_P$: pancreas mask).  
Our joint learning network architecture is shown in Fig.~\ref{fig: Framework}. This network has two branches for predicting both the resection margins and survival outcomes. The branch of resection margins uses one 3D-CNN model with six convolutional layers equipped with Batch Normalization and ReLu. Similar 3D architecture has shown good prediction performance for lung cancer~\cite{lou2019image}. Input of this branch is the concatenation of CT volumes at different time points and the corresponding tumor and pancreas masks: e.g., $X\in \mathbf{R}^{5 \times 64^3}$. This branch will try to learn the CT intensity attenuation variations and the relationships between tumor and surrounding pancreas regions, which help classify the tumor into different resection status. Note that R0/R1 can only be obtained after the surgery and pathology. Our model can be applied preoperatively in real scenarios to offer PDAC patients with the appropriate advice regarding surgical decisions.

The branch of predicting outcomes use CT volumes at each phase (each phase is CT-$M_T$-$M_P$ three-channel input, $X_t \in \mathbf{R}^{3\times64^3}$). The aim of this branch is to capture the tumor attenuation patterns across phases. Tumor attenuation usually means the contrast differences between the tumor and its surrounding pancreas regions so that we introduce both the tumor and pancreas masks into input volumes. The core part of this branch is a recurrence module that allows the network to retain what it has seen and to update the memory when it sees a new phase image. A naive approach is to use a vanilla LSTM or ConvLSTM network. Conventional ConvLSTM is capable of modeling 2D spatio-temporal image sequences by explicitly encoding the 2D spatial structures into the temporal domain~\cite{chen2016combining}. A more recent ST-ConvLSTM simultaneously learns both the spatial consistency among successive image slices and the temporal dynamics across different time points for the tumor growth prediction~\cite{zhang2019spatio}. Instead of using adjacent 2D CT slices and motivated by 3D object reconstruction~\cite{choy20163d},  we propose to use Contrast-Enhanced 3D Convolutional LSTM (CE-ConvLSTM) network to capture the temporally enhanced patterns from CE-CT sequences. CE-ConvLSTM can model 4D spatio-temporal CE-CT sequences by explicitly encoding their 3D spatial structures into the temporal domain. The main equations of ConvLSTM are as follows:
\begin{align}
    & f_t = \sigma(W_{f}^X * X_t + W_{f}^H*H_{t-1} + b_f) \\ \nonumber
    & i_t = \sigma(W_{i}^X*X_t + W_{i}^H*H_{t-1} + b_i) \\ \nonumber
    & o_t = \sigma(W_{o}^X*X_t + W_{o}^H*H_{t-1} + b_o) \\ \nonumber
    & C_t = f_t \odot C_{t-1} +i_t \odot tanh(W_{C}^X*X_t + W_{C}^H*H_{t-1} + b_{C}) \\ \nonumber
    & H_t = o_t \odot tanh(C_t)
\end{align}
where $X_t$ is the CE-CT sequences at time $t$, $*$ denotes the convolution operation, and $\odot$ denotes the Hadamard product. All the gates $f, i, o$, memory cell $C$, hidden state $H$ are 4D tensors. We use a $3 \times 3 \times 3$ convolutional kernel and 128 as the channel dimension of hidden states for the LSTM unit.
We employ 3D-ResNet18~\cite{hara2018can,chen2019med3d} as the encoder to encode each three-channel input to the lower-dimensional feature maps for CE-ConvLSTM. 

After the concatenation of feature maps from both tasks, the channel number of this common representation is 256. Then two separate fully-connected networks will use the common representation for each prediction task. In the training phase, labels of the resection status and patient overall survival information (OS time and censoring status) are known for each input CE-CT sequence. The weighted binary cross-entropy (BCE) loss is applied to the resection margin prediction task, while the negative log partial likelihood~\cite{katzman2018deepsurv,zhu2016deep} is used to predict the survival outcome $\mathbf{y}_i$ of this patient which is a continuous score and the loss function has been used in many deep learning survival models~\cite{yao2019deep,yao2020whole,Wulczyn2020}. This survival loss is summarized as
$L(\mathbf{y}_i) =\sum_{i}\delta_i(-\mathbf{y}_i +   \log{\sum_{j:t_j>=t_i}\exp(\mathbf{y}_j)})$
where $j$ is from the set whose survival time is equal or larger than $t_i$ ($t_j \ge t_i$).  

\section{Experiments}

{\bf Dataset.} Pancreatic CT scans of 205 patients (with resectable PDACs, mean tumor size = 2.5 cm) were undertaken preoperatively during non-contrast, pancreatic, and portal venous phases (i.e., 615 CT volumes). Only 24 out of 205 patients have R1 resection margins, and the imbalanced class weighting in the loss is considered. All images were resampled to an isotropic 1 mm$^3$ resolution. We adopt nested 5-fold cross-validation (with training, validation, and testing sets in each fold) to evaluate our model and other competing methods. 
To augment the training data, we rotate the volumetric tumors in the axial direction around the tumor center with the step size of $90^\circ$ to get the corresponding 3D CT image patches and their mirrored patches. We also randomly select the cropped regions with random shifts for each iteration during the training process. This data augmentation can improve the network's ability to locate the desired translational invariants. The batch sizes of our method and other models are the same as 8. The maximum iteration is set to be 500 epochs, and the model with the best performance on the validation set during training is selected for testing.

We first validate the performance from each branch shown in Fig.~\ref{fig: Framework} for single survival prediction task. Then we incorporate each model with CE-ConvLSTM and report the results as shown in Table \ref{tab:ci_CNN}. C-index value is adopted as our main evaluation metric for survival prediction. The C-index quantifies the ranking quality of rankings and is calculated as follows
$c=\frac{1}{n}\sum_{i\in\{1...N|\delta_i=1\}}\sum_{t_j>t_i}I[\mathbf{y}_i>\mathbf{y}_j]$
where $n$ is the number of comparable pairs and $I[.]$ is the indicator function. $t_i$ is the actual time observation of patient $i$ and $\mathbf{y}_i$ denotes the corresponding risk. The value of C-index ranges from 0 to 1. The averaged C-index using the single pancreatic phase is 0.659$\pm$0.075 from 3DCNN model. We can see that the prediction is improved when using early fused CE-CT images (CE-3DCNN, 0.662 vs. 0.659). When  CE-ConvLSTM is adopted by incorporating all phases, C-index is increased by 3\% (3DCNN-CE-ConvLSTM) compared with the model only using the pancreatic phase. 3D-ResNet18 with the pre-trained weights~\cite{chen2019med3d} is used as an advanced model compared to the conventional 3D ConvNets. 
From Table \ref{tab:ci_CNN}, ResNet3D with CE-ConvLSTM has improved performance versus ResNet3D with early fusion. Results in this table illustrate that the dynamic enhancement CT imaging patterns learned/captured by CE-ConvLSTM can help achieve significant prediction improvements compared against early fusion CNNs.

\begin{table}[t]\caption{Validation of CE-ConvLSTM with different CNN backbones.}
\begin{tabular}{l|c|c|c|c} \hline
Model   & Res-CE-ConvLSTM & CE-ResNet3D & 3DCNN-CE-ConvLSTM & CE-3DCNN  \\ \hline
C-index &     0.683$\pm$0.047 &  0.675$\pm$0.050 & 0.679$\pm$0.052 & 0.662$\pm$0.038 \\ \hline
\end{tabular}
\label{tab:ci_CNN}
\end{table}

\begin{table}[t]\caption{Results of different methods. Mul: multi-task; cls: single classification.}
\centering
\begin{tabular}{|c|c|c|c|c|c|c|}
\hline
\multirow{2}{*}{} & \multicolumn{1}{c|}{\multirow{2}{*}{Task}} & \multicolumn{2}{c|}{Survival}    & \multicolumn{3}{c|}{Resection Margin: R0/R1} \\ \cline{3-7} 
                  & \multicolumn{1}{c|}{}                            & \multicolumn{2}{c|}{C-index} & Balanced-ACC    & Sensitivity    & Specificity    \\ \hline
Proposed          & Mul                                              & \multicolumn{2}{l|}{\textbf{0.705$\pm$0.015}}  &   \textbf{0.736$\pm$0.141}    &    \textbf{0.813$\pm$0.222}     &  0.659$\pm$0.118       \\ \hline
Tang et al.\cite{tang2019pre}     & Mul                                             & \multicolumn{2}{l|}{0.683$\pm$0.056}        & 0.574$\pm$0.090 &  0.573$\pm$0.336      &  0.575$\pm$0.232       \\ \hline
Lou et al.\cite{lou2019image}     & Mul                                             & \multicolumn{2}{l|}{0.649$\pm$0.070}        & 0.682$\pm$0.091 &  0.673$\pm$0.236      &  0.690$\pm$0.080       \\ \hline
CE-ResNet3D          & cls                                            & \multicolumn{2}{c|}{-}        &  0.604$\pm$0.169      & 0.446$\pm$0.347        & \textbf{0.762$\pm$0.157}\\ \hline
CE-3DCNN             & cls                                            & \multicolumn{2}{c|}{-}        & 0.583$\pm$0.145       & 0.460$\pm$0.312  & 0.706$\pm$0.161         \\ \hline
\end{tabular}
\label{tab:resu_mul_task}
\end{table}
To further evaluate the performance of multi-task models, we report the results in comparison to recent multi-task deep prediction methods~\cite{tang2019pre,lou2019image}, as illustrated in Table \ref{tab:resu_mul_task}. Tang et al. propose using separate branches of 3D CNNs to predict both OS time and tumor genotype for glioblastoma (GBM) patients~\cite{tang2019pre}. Lou et al. present a multi-task training model on shared hidden representations from single model~\cite{lou2019image}.  We replace the RMSE loss from the original implementation of ~\cite{tang2019pre} with the negative log partial likelihood loss because it can handle alive patients (whereas authors discarded some patients who are still alive in~\cite{tang2019pre}). Those models cannot capture tumor temporal enhancement changes. Classification performances are evaluated by the metrics of Balanced-Accuracy, Sensitivity, and Specificity. Single classification task uses CE-ResNet3D and CE-3DCNN can be found in the last two rows in Table \ref{tab:resu_mul_task}. We can see the proposed method that uses CE-ConvLSTM to capture tumor enhancement patterns indeed achieves better performance in both tasks than the baseline multi-task models with early fusion.

\begin{table}[t]\caption{Univariate and Multivariate Cox regression analysis. HR: hazard ratio.}
\begin{tabular}{|l|c|c|c|c|c|}
\hline
& \multicolumn{3}{|l|}{Univariate Analysis} & \multicolumn{2}{l|}{Multivariate Cox}     \\ \hline
Factors              & HR (95\% CI)  & C-index  & p-value   & HR (95\% CI)               & p-value \\ \hline
Stromal Fraction     &  0.057(0.019-0.17)  & 0.659  &   6.38e-7 &    0.105(0.031-0.0.36)              & 3.39e-4       \\ \hline
Clear fat plane      &  3.771(2.235-6.364)  & 0.559   & 1.39e-5 &  2.010(1.157-3.490)  &0.013       \\ \hline
Tumor size           &   1.693(1.218-2.352)    &    0.588 & 0.002   &   1.568(1.108-2.22)     & 0.011       \\ \hline
Resection margin &  4.557(2.772-7.492)     &  0.577 &    2.99e-7 &3.704(2.147-6.390) &2.51e-6     \\ \hline
 TNM stage  &  1.475(1.107-1.966)     &    0.546    &    0.008  & 1.319(0.970-1.790)   &0.078  \\ \hline
 CA19.9     &  1.001(1.000-1.001)     &    0.557        &    0.0207  &   1.000(0.999-1.001)  &0.281    \\ \hline
Radiomics signature  & 1.181(1.109-1.257)  &  0.645  &    1.17e-4  & 1.085(1.004-1.171)&0.038         \\ \hline
Signature of~\cite{tang2019pre}  &  1.704(1.426-2.035)& 0.677 &  4.96e-9   &   1.035(0.824-1.30) & 0.767      \\ \hline
Our signature  &  1.764(1.478-2.105)& 0.691 &  2.81e-10   &   1.435(1.142-1.803) & 0.002      \\ \hline
\end{tabular}
\label{tab:stats} 
\end{table}

\begin{figure*}[t]
	\centering
	\includegraphics[width=1\linewidth]{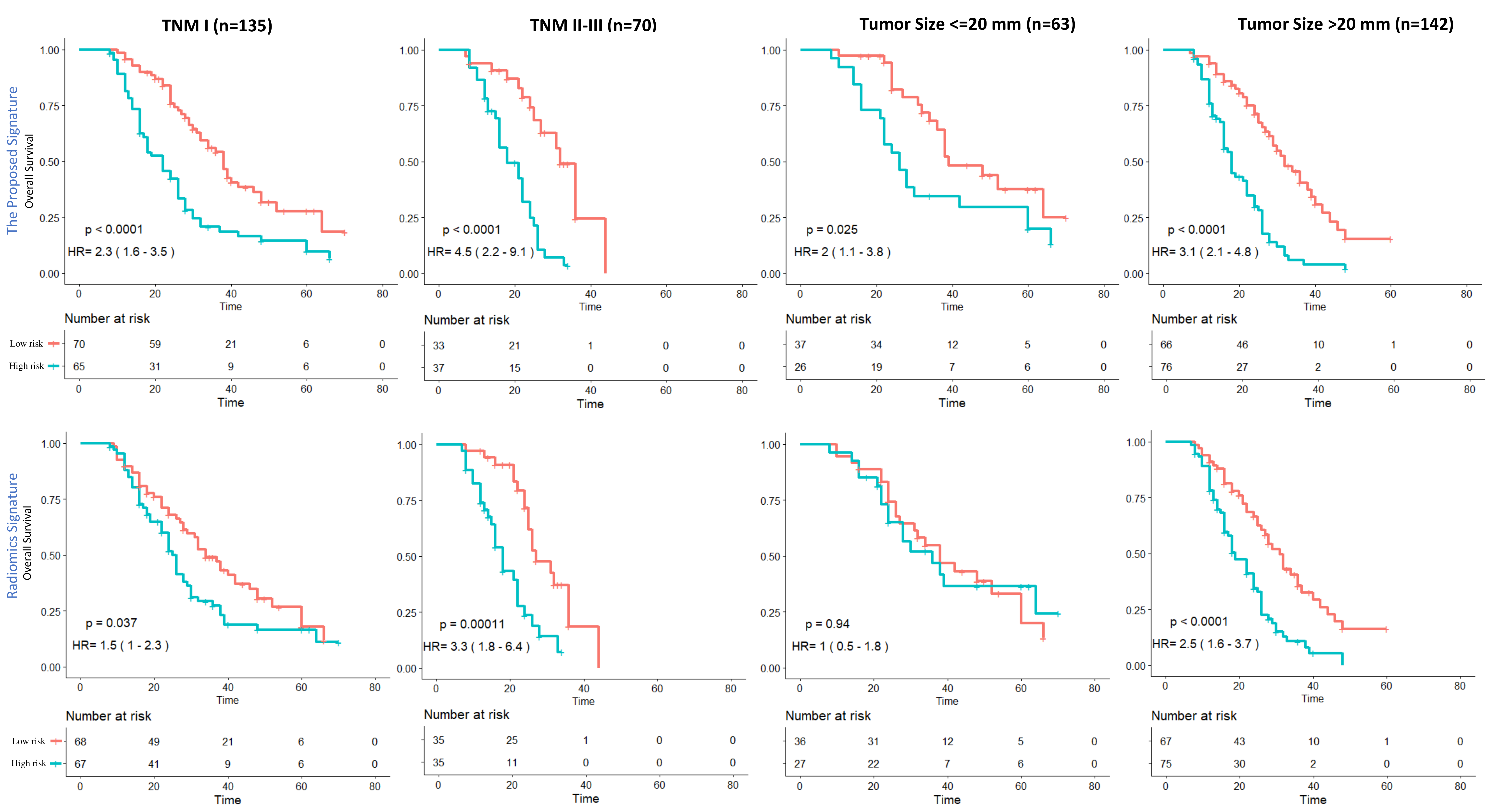}
	\caption{Kaplan-Meier analyses of overall survival according to the proposed (1st row) and radiomics signature (2nd row) in patients with two different stratifications (TNM staging and Tumor size). The proposed signature significantly stratifies all subgroups (TNM I vs II-III; or Tumor size $\le$ 20mm vs $>$20mm). 
	}
	\label{fig: kmplot}
\end{figure*}

In Table \ref{tab:stats}, univariate and multivariate cox proportional-hazards models are used to evaluate Hazard Ratio (HR) and log-rank test p-value for each factor including deep signature, radiomics signature and other clinicopathologic factors. 
To obtain the deep signature on all 205 patients, we use the mean and standard deviation of risk predictions on the training folds to normalize the risk scores of the corresponding testing fold. Then we combine results from each testing fold together. We also build radiomics signature using conventional radiologic features of input CE-CT images using Pyradiomics~\cite{van2017computational}. Each phase will provide 338 features, and thus each patient contains 1014 features in total. Features are then refined by Lasso-based Cox feature selection~\cite{tibshirani1997lasso}. Finally, the selected features are fed to the Cox regression model to get the signature. Each feature selection is performed on each training fold, and the number of selected features varies across different folds.
From the statistic analysis (Table \ref{tab:stats}), the proposed signature is the strongest prognostic factor in univariate analysis with the highest concordance index. The proposed signature remains strong in multivariable analysis  (HR=1.435, p=0.002) adjusting for established clinicopathologic prognostic markers, e.g.: stromal fractions (HR=0.105, $p<0.001$), resection margins (HR=3.704, $p<0.0001$), and TNM stage (HR=1.319, $p=0.078$). 
Our proposed one is even stronger than any other CT-derived signature using radiomics analysis or deep learning model without encoding tumor enhancement patterns.

To demonstrate the added value of the proposed signature to the current staging system, we plot Kaplan-Meier survival curves in Fig. \ref{fig: kmplot} for patients with further stratification by our signature after grouping by TNM and Tumor size, respectively, which are two well-established stratification criteria. The cut-off for our and radiomics signature is the median score of each signature. We study two subgroups of patients: 1) patients divided by TNM staging I vs. II-III, 2) patients divided by the primary PDAC tumor size $\le$ 20mm vs. $>$ 20 mm. It is shown that the proposed signature remains the most significant log-rank test outcome in the subgroup of patients, while the radiomics signature does not reach the statistical significance within the patient sub-population of PDAC tumor $\le$ 20mm. Results shown in Fig.~\ref{fig: kmplot} demonstrate that after using the current clinicopathologic TNM staging system or tumor size, our proposed multi-phase CT imaging based signature can indeed further provide the risk stratification with significant evidence. This described novel deep signature could be combined with the established clinicopathological criteria to refine the risk stratification and guide the individualized treatment of resectable PDAC patients.

\section{Conclusion}
In this paper, we propose a new multi-task CNN framework for cancer survival prediction by simultaneously predicting the tumor resection margins for resectable PDAC patients. The use of CE-ConvLSTM to consider and encode the dynamic tumor attenuation patterns of PDAC boost the whole framework, to significantly outperform the early fusion deep learning models and conventional radiomics-based survival models. Our results also validate that the proposed signature can serve as both a prognostic and predictive biomarker for subgroup patients after staging by the well-established pathological TNM or tumor size. This makes our model very promising in future clinical usage to refine the risk stratification and guide the surgery treatment of resectable PDAC patients. Future work will consider to integrate with other non-radiological factors (e.g. age, gender) to achieve better results.

\bibliography{paper213}
\bibliographystyle{splncs04}

\end{document}